\documentclass[a4paper]{jpconf}

\usepackage{graphicx}
\usepackage{bm}
\usepackage{slashed}

\begin{document}

\def\bb    #1{\hbox{\boldmath${#1}$}}

\title{Novel Bose-Einstein Interference in the Passage of a Fast
  Particle in a Dense Medium\footnote[1]{Based in part on an invited
    talk presented at the 35th Symposium on Nuclear Physics, Cocoyoc,
    Mexico, January 3, 2012.  } }

\author{Cheuk-Yin Wong} 

\address{
Physics Division, Oak Ridge National Laboratory, 
Oak Ridge, TN 37831
}

\ead{wongc@ornl.gov}

\begin{abstract}

When an energetic particle collides coherently with many medium
particles at high energies, the Bose-Einstein symmetry with respect to
the interchange of the exchanged virtual bosons leads to a destructive
interference of the Feynman amplitudes in most regions of the phase
space but a constructive interference in some other regions of the
phase space.  As a consequence, the recoiling medium particles have a
tendency to come out collectively along the direction of the incident
fast particle, each carrying a substantial fraction of the incident
longitudinal momentum.  Such an interference appearing as collective
recoils of scatterers along the incident particle direction may have
been observed in angular correlations of hadrons associated with a
high-$p_T$ trigger in high-energy AuAu collisions at RHIC.

\end{abstract}

\section{Introduction}

~~~~~In the collision of an energetic particle $p$=$(p_0,{\bb p})$
with $n$ medium particles, how dense must the medium be for the
multiple collisions to become a single coherent (1+$n$)-body
collision, instead of a sequence of $n$ incoherent 2-body collisions?
It is instructive to find out the conditions on the medium density and
the energy of the incident particle that determine whether the set of
multiple collisions are coherent or incoherent \cite{Won11a}.  For
such a purpose, we consider a binary collision between the incident
fast particle $p$ and a medium scatterer $a_i$ with the exchange of a
boson, in the medium center-of-momentum frame.  The longitudinal
momentum transfer $q_z$ for the binary collision can be obtained from
the transverse momentum transfer $q_T$ by $q_z \sim q_T^2/2p_0$.  The
longitudinal momentum transfer is associated with a longitudinal
coherence length $\Delta z_{\rm coh}\sim \hbar /q_z \sim 2\hbar
p_0/q_T^2$ that specifies the uncertainties in the longitudinal
locations at which the virtual boson is exchanged between the fast
particle and the scatterer.

The nature of the multiple scattering process can be inferred by
comparing the longitudinal coherence length $\Delta z_{\rm coh}$ with
the mean free path $\lambda$ of the jet in the dense medium that
depends not only on the density of the medium but also on the binary
collision cross section.  If $\Delta z_{\rm coh} \ll \lambda$, then a
single binary collision is well completed before another binary
collision begins, and the multiple collision process consists of a
sequence of $n$ incoherent two-body collisions.

If $\Delta z_{\rm coh} \gg \lambda$, then a single binary collision is
not completed before another one begins, and the multiple collision
process consists of a set of coherent collisions as a single
(1+$n$)-body collision.  For a set of initial and final states in such
a coherent (1+$n$)-body collision, there are $n!$ different
trajectories in the sequence of collisions along $\Delta z_{\rm coh}$
at which various virtual bosons are exchanged.  By Bose-Einstein
symmetry, the total Feynman amplitude is then the sum of the
amplitudes for all possible interchanges of the exchanged virtual
bosons.  The summation of these Feynman amplitudes and the
accompanying interference constitute the Bose-Einstein interference in
the passage of the fast particle in the dense medium.

There is another important effect that accompanies the coherent
(1+$n$)-body collision and changes the nature of the collision
process.  A sequence of $n$ incoherent two-body collisions contains
only $2n$ degrees of freedom, which can be chosen to be the transverse
momentum transfers $\{ {\bb q}_{1T},{\bb q}_{2T},{\bb q}_{3T},...,{\bb
  q}_{nT}\}$.  The longitudinal momentum transfer $q_z$ after each
individual jet-parton collision is then a dependent variable,
depending on the corresponding transverse momentum transfer as $q_{iz}
\sim |\bb q_{iT}|^2/2p_0$.  In contrast, the coherent (1+$n$)-body
collision links the incident particle with $n$ scatterers as a single
collisional unit.  There are $3n-1$ degrees of freedom in this
coherent (1+$n$)-body collision, after the $3(n+1)$ degrees of freedom
are reduced by the constraints of the conservation of energy and
momentum.  The set of longitudinal momentum transfers, $\{ q_{iz},
i=1,2,...,n\}$, can also be independent variables with their own
probability distribution functions.  The longitudinal momentum of the
incident fast particle can flow to the $n$ scatterers, which can then
carry a substantial fraction of the longitudinal momentum of the
incident fast particle.  Thus, the degrees of freedom increases from
$2n$ for incoherent collisions to $3n-1$ for a coherent collision, and
this increase changes the character of the collision process.

It is of interest to inquire whether there are fast particles
energetic enough, scatterer media dense enough, and binary cross
section large enough, for coherent collisions to occur.  In high
energy central collisions between heavy nuclei such as those at RHIC
and LHC, both jets (mini-jets) and a dense medium are produced after
each collision.  The jets will collide with partons in the dense
medium, and these collisions may satisfy the condition for coherent
collisions.  The longitudinal coherent length $\Delta z_{\rm coh}$ is
of order 25 fm, for a typical transverse momentum transfer of
$q_T$$\sim$0.4 GeV/c from a jet of momentum $p_0$$\sim$10 GeV/c to a
medium parton in a binary collision at RHIC \cite{Ada08}.  The
longitudinal coherent length $\Delta z_{\rm coh}$ is much greater than
the radius $R$ of a large nucleus.  On the other hand, the away side
jet is quenched by the dense medium in the most central AuAu
collisions at RHIC and LHC \cite{Bjo82,Gyu94,Adl04}, and the near-side
jet collides with about 4-6 medium partons \cite{Won07}-\cite{Won11}.
Therefore, one can infer that the mean-free path $\lambda$ for the
collision of the jet with medium partons is much smaller than the
nuclear radius $R$.  Hence, $\Delta z_{\rm coh} \gg R \gg \lambda$ in
high-energy central nuclear collisions at RHIC and LHC, and the
multiple collision process of a jet with medium partons along its path
constitutes a set of coherent collision.

\section{Bose-Einstein Interference of Feynman Amplitudes}

~~~As an example of the  interference of 
Feynman amplitudes in a coherent collision, we consider 
\begin{eqnarray}
p+a_1+a_2 \to p'+a_1'+a_2',
\label{1}
\end{eqnarray}
the collisions of
an energetic fermion $p$ with two fermion scatterers $a_1$ and $a_2$ of rest mass $m$ in the
Abelian gauge theory.  
\begin{figure} [h]
\hspace*{4.0cm}
\includegraphics[scale=0.45]{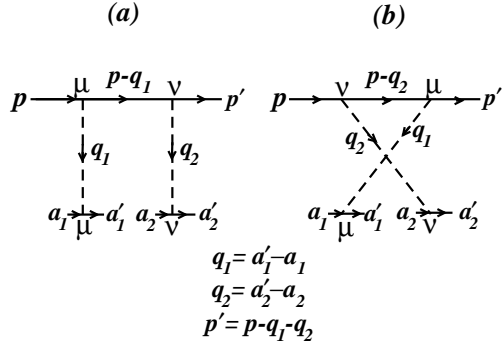}
\caption{ Feynman diagrams for the collision of a fast fermion $p$
  with medium fermions $a_1$, $a_2$, with the emission and absorption
  of virtual bosons of momenta $q_1$ and $q_2$.  }
\end{figure}
The Feynman  amplitude for diagram 1(a) is
given by \cite{Che87}
\begin{eqnarray}
M_a &= &- g^4 {\bar u}({\bb p}')
  \gamma_\nu \frac{1}{\slashed p - \slashed q_1-m+ i \epsilon'} 
\gamma_\mu u({\bb p})
\frac{1}{q_2^2}  {\bar u}({\bb a}_2')
\gamma_\nu u({\bb a}_2)
\frac{1}{q_1^2}  {\bar u}({\bb a_1}')
\gamma_\mu u({\bb a}_1).
\end{eqnarray}

If the spatial separation between the scatterers is so large that
$\lambda \gg \Delta z_{\rm coh}$, then diagram 1(a) can be cut into
two disjoint pieces and the collision process consists effectively of
a set of two incoherent two-body collisions.

On the other hand, if the collision process is characterized by
$\lambda \ll \Delta z_{\rm coh}$, it comprises a set of coherent
(1+2)-body collision.  There is an additional Feynman amplitude $M_b$
for diagram 1(b), obtained by making a symmetrized permutation of the
bosons in diagram 1(a),
\begin{eqnarray}
M_b &= &- g^4 {\bar u}({\bb p}')
  \gamma_\mu \frac{1}{\slashed p - \slashed q_2-m+ i \epsilon'} 
\gamma_\nu u({\bb p})
\frac{1}{q_2^2}  {\bar u}({\bb a}_2')
\gamma_\nu u({\bb a}_2)
\frac{1}{q_1^2}  {\bar u}({\bb a_1}')
\gamma_\mu u({\bb a}_1).
\end{eqnarray}
The trajectories for diagram 1(a) and 1(b) are both possible paths in
a coherent collision, leading from the a set of initial states to a
set of final states.  By Bose-Einstein symmetry, the total amplitude
$M$ for coherent collisions is the symmetrized sum of $M_a$ and $M_b$.

We  consider the high-energy limit and assume the conservation of helicity with 
\begin{eqnarray}
\{ p_0, |{\bb p}| ,p_0', |{\bb p}'|\} \gg \{ |{\bb a_i }|, q_0,|{\bb q_i }|\}  \gg m, {\rm ~for~}
i=1,2.\label{2a}
\end{eqnarray}
In this limit, we have approximately \cite{Che87}
 \begin{eqnarray}
&&{\bar u}({\bb a}') \gamma_\mu u({\bb a})
\sim
\sqrt{\frac{a_0+m}{{a_0'+m} }}
\frac{a_\mu'}{2m}
+ \sqrt{\frac{a_0'+m}{a_0+m} }
\frac{a_\mu}{2m} \equiv  \frac {{\tilde a}_\mu}{m},\label{5} \\
&&\frac{1}{\slashed p - \slashed q_1-m +i \epsilon'} 
\gamma_\nu u(p)
\sim  - \frac{\slashed p - \slashed q_1+m }{2p \cdot q_1 - i \epsilon} 
\gamma_\nu u(p),\label{5}
\\
&&{\bar u}(p) \gamma_\nu  (\slashed p - \slashed q_1+m)   \gamma_\mu u(p)
\sim   \frac{ 2 p_{\nu}p_{\mu}}{m}, 
\end{eqnarray}
where $\epsilon$ is a small positive quantity.  We shall be interested
in the case in which the fermion $p'$ after the collision remains on
the mass shell.  The mass shell condition can be expressed as
 \begin{eqnarray}
(p-q_1-q_2)^2-m^2~\sim~ -2p\cdot q_1-2p\cdot q_2 ~\sim~ 0.
\label{6}
\end{eqnarray}
The symmetrized sum of the Feynman amplitudes
$M_a$ and $M_b$ in the high-energy limit is
\begin{eqnarray}
M
\sim
\frac{g^4}{2m}  \frac{2 p\cdot {\tilde a}_1}{m}
\frac{ 2 p\cdot {\tilde a}_2}{m}
\frac{1}{q_2^2 q_1^2}
\left (
 \frac{1}{2 p \cdot  q_1 - i \epsilon} 
+\frac{1}{2 p \cdot  q_2 - i \epsilon}
\right ).
\label{11}
\end{eqnarray} 
Note that the amplitudes $M_a$ and $M_b$ correlate with each other
because of the mass-shell condition (\ref{6}).  The real parts of the
amplitudes destructively cancel, and the imaginary parts interfere and
add constructively, to result in sharp distributions at $p \cdot q_1$
$\sim$ 0 and $p \cdot q_2$ $\sim$ 0,
\begin{eqnarray}
M
\sim
\frac{g^4}{2m}  \frac{2 p\cdot {\tilde a}_1}{m}
\frac{ 2 p\cdot {\tilde a}_2}{m}
\frac{1}{q_2^2 q_1^2}
\biggl  \{
 i \pi \Delta(2 p \cdot  q_1)+i\pi \Delta(2 p \cdot  q_2)
\biggr \},
~~~\Delta(2p\cdot q_1)=\frac{1}{\pi}\frac{\epsilon}
{(2 p \cdot  q_1)^2+\epsilon^2},
\end{eqnarray} 
where the function $\Delta(2p\cdot q_1)$ approaches the Dirac delta function
$\delta(2p\cdot q_1)$ in the limit $\epsilon \to 0$.  

Generalizing to the case of the coherent collision of a fast fermion
with $n$ fermion scatterers, the total Feynman amplitude is
\begin{eqnarray}
M=\frac{ g^{2n}}{2m} 
\left \{  \prod_{i=1}^n \frac{2 p\cdot {\tilde a}_i }{mq_i^2 } \right \}
{\cal M}(q_1,q_2,...,q_n),
\label{10}
\end{eqnarray}
where  ${\cal M}(q_1,q_2,...,q_n)$ is the
sum of  $n!$ amplitudes involving symmetric
permutations of the exchanged bosons given by
\begin{eqnarray}
{\cal M}(q_1,q_2,...,q_n)=\prod_{j=1}^{n-1}\frac{1}
{\sum_{i=1}^j 2p\cdot q_i - i \epsilon}+ {\rm symmetric~ permutations}.
\end{eqnarray}
The above sum 
involves extensive cancellations.
Remarkably, it can be shown that this sum 
of $n!$ permutations
turns out to be a product of sharp distributions centered at $2p \cdot q_i \sim 0$
\cite{Che87},
\begin{eqnarray}
{\cal M}(q_1,q_2,...,q_n)\Delta (\sum_{i=1}^n 2p\cdot q_i)= 
(2\pi i)^{n-1}\prod_{i=1}^n   \Delta (2p\cdot q_i),
\label{13}
\end{eqnarray}
which indicates that in coherent collisions, there is a destructive
interference of the Feynman amplitudes in most regions of the phase
space but a constructive interference in some other regions of the the
phase space, leading to sharp distributions at $2p \cdot q_i \sim 0$.

Equations (\ref{10})-(\ref{13}) for the total Feynman amplitude as a
product of delta functions of $2p\cdot q_i$ for a coherent collision
are similar to previous results obtained for the emission of many real
photons or gluons in bremsstrahlung, and for the sum of ladder and
cross-ladder amplitudes in the collision of two fermions
\cite{Che87}-\cite{Lam97b}.

\section{Consequences of the BE Interference on the Recoils of Fermion Scatterers}
~~~The differential cross section for the coherent collision
$p+a_1+...+a_n \to p' + a_1'+ ...+a_n'$ is
\begin{eqnarray}
d^{n}\sigma = \frac{1}{f_{j1}f_{j2}...f_{jn}T_2 T_3...T_n} {|M|^2}   (2\pi)^4 \delta^4 
(p'+\sum_{i=1}^n q_i -p)\frac{d{\bb p}_i'2m}{(2\pi)^3 2 p_{i0}' }{\prod_{i=1}^n   \frac{d^4 a_i'2m}{(2\pi)^3 2 a_{i0}' }D_i( a_{i}')},
\end{eqnarray}
where $f_{ji}$ is the dimensionless flux factor between the fast
particle $p$ and scatterer $a_i$,
\begin{eqnarray}
f_{ji}&=&\frac{\sqrt{ p\cdot a_{i} -m ~m_{iT}}}{m_i ~ m},
\end{eqnarray}
and $\{T_i, i=1,..(n-1)\}$ are the collision times for the
intermediate state of the fast particle $p$ to collide with the
scatterers after the $i$th collision, and $m_{iT}=\sqrt{m_i^2+{\bb
    a}_T^2}$.  The function $D_i$ describes the mass-shell conditions
of medium scatterers after collision,
\begin{eqnarray}
D_i( a_{i}') =\frac{\Gamma_i/2}{\pi\{  [(a_{i0}'-
\sqrt{(\bb a_i' - \bb A)^2 + (m_i+S)^2 } + A_0]^2 + \Gamma_i^2/4 \}},
\end{eqnarray}
where $A=\{\bb A, A_0\}$ and $S$ are the vector and scalar mean fields
experienced by the medium scatterer $a_i'$ after collision,
respectively, and $\Gamma_i$ is the width of $a_i'$.  As the mean
fields and scatterer widths increase with density and are presumably
quite large and dominant for a dense medium, we approximately
represent $D_i$ as an average constant.  We change variables from $
a_i'$ to $ q_i=a_i'-a_i$, and integrate over ${\bb p}'$.  Using
Eq.\ (\ref{13}), we obtain
\begin{eqnarray}
d^{n} \sigma
=
  \frac{g^{4n}(2m)^{n-1}}{(2\pi)^{n+1} }
\left \{ \prod_{i=1}^{n}\frac{D_i}{f_{ij}} \right \}
\frac{
\left \{\prod_{i=1}^{n-1} 
\Delta(2 p\cdot q_i)\right \}}
{T_1T_2 T_3 T_4...T_{n-1}}
\left \{\prod_{i=1}^{n} 
\Delta(2 p\cdot q_i)dq_{i0}\right \}
\left \{ \prod_{i=1}^n
 \frac{(2 p \cdot {\tilde a}_i)^2 }{m^2 q_i^4}
\frac{d \bb q_{iT} dq_{iz} }{ 2 a_{i0}' }\right \}.
\label{18}
\end{eqnarray}
The distribution
$\Delta(p \cdot q_i)$ can be written as
\begin{eqnarray}
\Delta (2p\cdot q_i)=\frac{1}{p_0+p_z}
\Delta \biggl  (q_{i0}-q_{iz} - \frac{ -(p_{0}-p_z)( q_{i0}+q_{iz}) + 2{\bb p}_T\cdot\bb q_T}{p_0+p_z} 
\biggr ),
\end{eqnarray}
which provides the constraint
$ q_{i0}-q_{iz}\sim 0.$
As a consequence, the integral of
$ |\Delta (2p \cdot q_i)|^2 dq_{i0}$ can be carried out, with the collision time $T_i$ canceling one of the factors in
$\Delta (2p \cdot q_i)|_{2p \cdot q_i\to 0}$.
The denominators of the boson propagator in
Eq.\ (\ref{18}) becomes
\begin{eqnarray}
q_i^2=(q_{i0}+q_{iz}) ( q_{i0}-q_{iz})   - |{\bb q}_{iT}|^2\approx - |{\bb q}_{iT}|^2 .
\end{eqnarray}
We obtain 
\begin{eqnarray}
d^n \sigma &=&
\frac{\alpha^2}{p_z m}
\left ( \frac{\alpha^2}{p_z^2}\right )^{n-1}
\left \{ \prod_{i=1}^{n}\frac{[2mD_{i}]}{f_{ij}} \right \} 
\left \{ \prod_{i=1}^n \frac{ (2p\cdot {\tilde a}_i)^2  ~ dq_{iz} d{\bb q}_{iT}}{  m^2  2a_{i0}' |\bb q_T|^4}\right \},
\label{22}
\end{eqnarray}
where $\alpha$=$g^2/4\pi$, the last two factors are dimensionless, and
$d^n\sigma$ has the dimension of $(\alpha^2/p_z^2)^n$.

To find the probability distribution for the longitudinal momentum
transfer $q_{iz}$, we introduce the fractional longitudinal momentum
kick
\begin{eqnarray}
x_i=\frac{q_{iz}}{p_z},~~~~~
dq_{iz}=p_z dx_i.
\end{eqnarray}
To investigate the $x_i$ dependence of the factor  $(2p\cdot
{\tilde a_i})^2/{2a_{i0}'}$ in Eq.\ (\ref{22}), we note from
Eq.\ (\ref{5}) that ${\tilde a_i}$ can be written as a function of
$q_i$ and $a_i$,
\begin{eqnarray}
\tilde a_i \sim \sqrt{\frac{a_{i0}+m}{{a_{i0}'+m} }}\frac{q_i}{2}
+\frac{a_{i0}'+m + a_{i0}+m}{\sqrt{(a_{i0}'+m)(a_{i0}+m) }}
\frac{a_i}{2}.
\end{eqnarray}
Because of the  $\Delta(2p\cdot q_i)$ constraint,
the factor $(2p\cdot {\tilde a_i})^2/{2a_{i0}'}$  in Eq.\ (\ref{22}) 
becomes 
\begin{eqnarray}
\frac{(2p\cdot {\tilde a_i})^2}{2a_{i0}'}
 \sim  \frac{(a_{i0}'+ a_{i0})^2}{2(a_{i0}')^2 }
\frac{(p\cdot {a_i})^2}{a_{i0}}
\equiv \kappa_i \frac{(p\cdot {a_i})^2}{a_{i0}}.
\end{eqnarray}
We obtain from Eq.\ (\ref{25})
\begin{eqnarray}
d^n \sigma 
=
\left \{
\frac{\alpha^2}{p_z m}
\left ( \frac{\alpha^2}{p_z^2}\right )^{n-1}
\frac{p_z^{n}}{m^{2n}}
\left ( \prod_{i=1}^{n}\frac{[2mD_{i}]}{f_{ij}} 
 \frac{\kappa_i (p\cdot a_i)^2}{a_{i0}} \right )\right \}
\frac{dx_1 dx_2 ... dx_n d{\bb q}_{1T}d{\bb q}_{2T}...d{\bb q}_{nT}}
{  |\bb q_{1T}|^4~|\bb q_{2T}|^4~...~|\bb q_{nT}|^4}.
\label{25}
\end{eqnarray}
The fermion scatterers can possess different initial energies $
a_{i0}$ at the moment of their collisions with the energetic jet.  In
the case when $ a_{i0} \ll q_{i0}$, the factor $\kappa_i$ approaches
$1/2+O(a_{i0}/q_{iz})$ with $(a_{i0}/q_{iz}) \ll 1$.  In the other
extreme when $ a_{i0} \gg q_{i0} $, the factor $\kappa$ approaches $2
+ O(q_{iz}/a_{i0})$ with $(q_{iz}/a_{i0}) \ll 1$.  The dependence of
$\kappa_i$ on $x_i$ is weak in either limits and can be neglected in
our approximate estimate.  We obtain then approximately
\begin{eqnarray}
d^n \sigma 
\sim 
\biggl  \{ {\rm constant~factor} \biggr  \}~
 \frac{dx_1 dx_2 ... dx_n ~~
d{\bb q}_{1T}  d{\bb q}_{2T} ...d{\bb q}_{nT} }
{  |\bb q_{1T}|^4~|\bb q_{2T}|^4~...~|\bb q_{nT}|^4},
\label{25}
\end{eqnarray}
where the constant factor is approximately independent of $x_i$ and
${\bb q}_{iT}$.  By symmetry, the fraction of momentum transfers $x_i$
for different scatterers should be approximately the same on the
average, and $x_i^{\rm max} $$\sim$ $1/n$.  Then as far as $x_i$ is
concerned, the average distribution is
\begin{eqnarray}
\frac{dP}{dx_i}\biggr |_{x_j=x_i,j=1,...,n} \sim  n~ \Theta(\frac{1}{n}-x_i),
\end{eqnarray}
and the average longitudinal momentum fraction 
is
\begin{eqnarray}
\langle x_i\rangle \sim \frac{1}{2n}~~~{\rm or}~~\langle q_{iz}\rangle \sim \frac{p_z}{2n}.
\label{27}
\end{eqnarray}
The above results in Eqs.\ (\ref{25})-(\ref{27}) indicate that in the
passage of an energetic fermion making coherent collisions with medium
partons, the reaction has a high probability for the occurrence of
small values of $ |{\bb q}_{iT}|$.  The singularities at $|\bb q_{iT}|
\sim 0$ in Eq.\ (\ref{25}) correspond to the case of infrared
instabilities that may be renormalized, and a momentum cut-off
$\Lambda_{\rm cut}$ may be introduced.  The scatterers acquire an
average longitudinal momentum $\langle q_{iz}\rangle \sim p_z/2n$ that
is expected to be much greater than $\langle |\bb q_{iT}|\rangle$.
Thus, there is a collective quantum many-body effect arising from
Bose-Einstein interference such that the fermion scatterers emerge in
the direction of the incident particle, each carrying a fraction of
the forward longitudinal momentum of the incident particle that is
inversely proportional to twice the number of scatterers, $\langle
q_{iz}\rangle \sim {p_z}/{2n}$.

\section{Bose-Einstein Interference for  Coherent Collisions  in Non-Abelian Theory}

~~~The above considerations for the Abelian theory can be extended to
the non-Abelian theory.  As an example, we consider a quark jet $p$
making coherent collisions with quarks $a_1$ and $a_2$ in the reaction
$p+a_1+a_2 \to p'+a_1'+a_2'$, in the non-Abelian theory.  We shall
neglect four-particle vertices and loops, which are of higher-orders.
The Feynman diagrams are then the same as those in Fig.\ 1.  One
associates each quark vertex with a color matrix
$T_{\alpha,\beta}^{(p,1,2)}$ where the superscript $p,1,$ or 2
identifies the quark $p$, $a_1$, or $a_2$, and the subscripts $\alpha$
or $\beta$ give the $SU(3)$ color matrix index.  The Feynman amplitude
$M_a$ for diagram 1(a) is
\begin{eqnarray}
M_a \!=\! -g^4 {\bar u}({\bb p}')
 T_\beta^{(p)} \gamma_\nu \frac{1}{\slashed p - \slashed q_1-m+ i \epsilon'} 
T_\alpha^{(p)}\gamma_\mu u({\bb p})
\frac{1}{q_2^2}  {\bar u}({\bb a}_2')
T_\beta^{(2)} \gamma_\nu u({\bb a}_2)
\frac{1}{q_1^2}  {\bar u}({\bb a_1}')
T_\alpha^{(1)}\gamma_\mu u({\bb a}_1).
\nonumber
\end{eqnarray}
The Feynman  amplitude $M_b$ for  diagram 1(b) is
\begin{eqnarray}
M_b\! =\! -g^4 {\bar u}({\bb p}')
T_\alpha^{(p)}  \gamma_\mu \frac{1}{\slashed p - \slashed q_2-m+ i \epsilon'} 
T_\beta^{(p)} \gamma_\nu u({\bb p})
\frac{1}{q_2^2}  {\bar u}({\bb a}_2')
T_\beta^{(2)} \gamma_\nu u({\bb a}_2)
\frac{1}{q_1^2}  {\bar u}({\bb a_1}')
T_\alpha^{(1)}\gamma_\mu u({\bb a}_1).\nonumber
\end{eqnarray}
In the high-energy limit for coherent collisions, the sum of the Feynman amplitudes is
\begin{eqnarray}
&&M
\sim
\frac{g^4}{2m}  \frac{2 p\cdot {\tilde a}_1}{m}
\frac{ 2 p\cdot {\tilde a}_2}{m}
\frac{1}{q_2^2 q_1^2}
\left (\frac{T_\beta^{(p)} T_\alpha^{(p)}T_\beta^{(2)} T_\alpha^{(1)}}{2p\cdot q_1 -i\epsilon}
+\frac{ T_\alpha^{(p)} T_\beta^{(p)}T_\beta^{(2)} T_\alpha^{(1)}}{2p\cdot q_2 -i\epsilon}
\right ).
\label{44}
\end{eqnarray}
We can rewrite the product of the color matrices for the  quark jet $p$ as 
\begin{eqnarray}
T_\beta^{(p)} T_\alpha^{(p)}=\frac{1}{2}\left ( [ T_\beta^{(p)}, T_\alpha^{(p)}]_+ + [ T_\beta^{(p)}, T_\alpha^{(p)}]_- \right ), \\
T_\alpha^{(p)} T_\beta^{(p)}=\frac{1}{2}\left ( [ T_\beta^{(p)}, T_\alpha^{(p)}]_+ - [ T_\beta^{(p)}, T_\alpha^{(p)}]_- \right ) .
\end{eqnarray}
For coherent collisions, the Feynman amplitude is then
 \begin{eqnarray}
&M
\sim
\frac{g^4}{2m}  \frac{2 p\cdot {\tilde a}_1}{m}
\frac{ 2 p\cdot {\tilde a}_2}{m}
\frac{1}{q_2^2 q_1^2}
\biggl \{ ({\cal M}_1+{\cal M}_2) 
\frac{[T_\beta^{(p)},T_\alpha^{(p)}]_+T_\beta^{(2)} T_\alpha^{(1)}}{2}
+
 ({\cal M}_1-{\cal M}_2) 
\frac{[T_\beta^{(p)},T_\alpha^{(p)}]_- T_\beta^{(2)} T_\alpha^{(1)}}{2}
 \biggr \}, \label{49}\\
&{\cal M}_1  + {\cal M}_2
=i \Delta(2 p \cdot  q_1)+i\Delta(2 p \cdot  q_2),
~~~{\rm and}~~~
{\cal M}_1 - {\cal M}_2  =
\frac{4 p \cdot  q_1}
{(2 p \cdot  q_1)^2+\epsilon^2}.
\end{eqnarray}

From the above analysis, we find that the color degrees of freedom in
QCD bring in additional properties to the Feynman amplitudes.
Bose-Einstein symmetry with respect to the interchange of gluons in
QCD involves not only the space-time exchange symmetry but also color
index exchange symmetry.  The total exchange symmetry can be attained
with symmetric space-time amplitudes and symmetric color index factors
as in the $({\cal M}_1 + {\cal M}_2)$ term in Eq.\ (\ref{49}).
The total symmetry can also be attained with space-time antisymmetry
and color index antisymmetry, as in the $({\cal M}_1-{\cal M}_2)$
term in Eq.\ (\ref{49}) .

For the space-time symmetric and color index exchange symmetric
component, the Feynman amplitude is equal to the Abelian Feynman
amplitude multiplied by a color factor. It will exhibit the same
degree of Bose-Einstein interference as in the Abelian theory.
Previous analysis on the longitudinal momentum transfer of recoiling
fermions in the Abelian theory  can be applied to the
non-Abelian theory for this space-time symmetric and color index
exchange symmetric component.  There is thus a finite probability for
the presence of  constraints to lead to recoiling
quarks receiving significant momentum kicks along the direction of the
incident quark jet.

The above considerations for jets and scatterers can be extended to cases involving gluon jets and gluon scatterers.  The
results are similar but with a small modification on the distribution for gluon scatterers  \cite{Won11a}.

\section{Signatures  for Bose-Einstein Interference in the Passage of a Jet }

~~~~From the results in the above sections, the occurrence of the
Bose-Einstein interference in the coherent collisions of a jet with
medium partons possesses the following characteristics:

\begin{itemize}
\item[(1)]
The Bose-Einstein interference is a quantum many-body effect.  It
occurs only in the coherent multiple collisions of the fast jet with two or
more scatterers,
$n\ge 2$.

\item[(2)]
Each scatterer has a transverse momentum distribution 
of the type $1/|{\bb q}_T|^4$, which peaks at small values of $|{\bb q}_T|$.

\item[(3)]
Each scatterer acquires a longitudinal momentum kick $q_z$ along the
incident jet direction that is approximately inversely proportional to twice
the number of scatterers, with $\langle q_{z}\rangle\sim p_z/2n$.

\item[(4)]
As a consequence, the final effect is the occurrence of collective
recoils of the scatterers along the jet direction.
\end{itemize}

\begin{figure} [h]
\hspace*{5.0cm}
\includegraphics[scale=0.50]{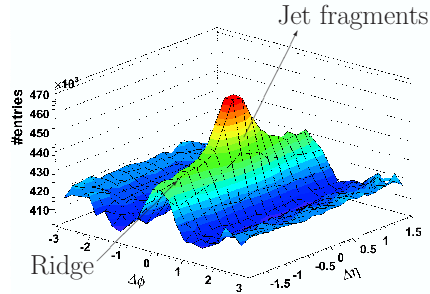}
\caption{$\Delta \phi$-$\Delta \eta$ correlation of produced hadrons  in the most central AuAu collision at RHIC with a high $p_T$ trigger,  measured by the STAR Collaboration \cite{Put07}.  }
\end{figure}

In a high energy central collision of heavy nuclei, both jets and a
dense medium are produced soon after the collision and jets propagate
in the dense medium.  The mean free path for the jet in the medium may
be small compared with the longitudinal coherent length to provide the
appropriate environment for coherent collisions.  It is of interest to
search for evidence for Bose-Einstein interference in the passages of
the jet in the dense medium.  In such a search, we need to separate
out jet fragments from recoiling scatterers among the detected
hadrons.  Such a separation is indeed kinematically possible in the
$\Delta \phi$-$\Delta\eta$ correlation measurements where $\Delta
\phi=\phi({\rm associated})-\phi({\rm trigger})$ is the azimuthal
angle difference between the produced hadron pair, and
$\Delta\eta=\eta({\rm associated})-\eta({\rm trigger})$ is the
pseudorapidity difference.  We show in Fig.\ 2 the $\Delta
\phi$-$\Delta\eta$ correlation data from the STAR Collaboration
\cite{Put07} for the most central AuAu collision at RHIC with a high
$p_T$ trigger \cite{Put07}.  Jet fragments are distributed in a small
cone of $(\Delta \phi, \Delta \eta)$$\sim$ 0 \cite{Ada05}-\cite{
  Lee09}.  Medium particles are distributed in the ``ridge" part at
$\Delta \phi \sim 0$ (Fig. 2) along the $\Delta \eta$ axis.  In
particular, ridge particle with $|\Delta \eta| > 0.6$ and $\Delta \phi
\sim 0$ along the ridge (Fig. 2) can be identified as medium scatterer
partons for the following reasons:

\begin{enumerate}

\item
The yield of the ridge particles increases approximately linearly with
the number of participants \cite{Put07}.

\item
The yield of the ridge particles is nearly independent of (i) the
flavor content, (ii) the meson/hyperon character, and (iii) the
transverse momentum $p_T$ (above 4 GeV) of the jet trigger
\cite{Put07,Bie07}.

\item
The ridge particles have a temperature (inverse slope) that is similar
(but slightly higher) than that of the inclusive bulk particles, but
lower than the temperature of the near-side jet fragments \cite{Put07}

\item
The baryon/meson ratio of the ridge particles is similar to those of
the bulk hadrons and is quite different from those in the jet
fragments \cite{Lee09}.

\item
The Wigner function analysis indicates that the momentum distribution
of medium partons has a rapidity plateau structure even at the early
stage of flux tube fragmentation \cite{Won09a}.

\end{enumerate}

With the medium scatterers (ridge particles) separated from the
incident high-$p_T$ jet fragments, the occurrence of the Bose-Einstein
interference can be characterized by the collective recoils of the
scatterers (the ridge particles) along the jet direction.  The
collective recoils will lead to the $\Delta \phi \sim 0$ correlation
of the ridge particles with the high-$p_T$ trigger.  As shown in
Fig.\ 2, such collective recoils of the medium particles have indeed
been observed in $\Delta \phi$-$\Delta \eta$ correlations of produced
hadrons in AuAu collisions at RHIC by the STAR Collaboration
\cite{Ada05}-\cite{Lee09}.  Similar $\Delta \phi$-$\Delta \eta$
correlations have also been observed by the PHENIX Collaboration
\cite{Ada08,Mcc08,Jia08qm}, and the PHOBOS Collaboration \cite{Wen08}.
Quantitatively, the collective recoils of the kicked medium partons
have been encoded into the longitudinal momentum kick $\langle q_{z}
\rangle$ of the momentum kick model analysis that yields the observed
$\Delta \phi$, $\Delta \eta$, and $p_T$ dependencies of the angular
correlations \cite{Won07}-\cite{Won11}.

It is of interest to examine the relationship $\langle q_{z}\rangle
\sim {p_z}/{2n}$ between $n$ and $\langle q_{z} \rangle$, when such a
collective momentum kick occurs.  For the most central AuAu collisions
at $\sqrt{s_{NN}}=200$ GeV at RHIC, we previously found from the
momentum kick model analysis that $ \langle n \rangle \sim 6$ and $q_z
\sim 1$ GeV \cite{Won08}.  In another momentum kick model analysis for
the highest multiplicity $pp$ collisions at $\sqrt{s_{NN}}=7$ TeV at
the LHC, we previously found that $\langle n \rangle \sim 2.4$ and
$q_z\sim 2 $ GeV.  The experimental data give a longitudinal momentum
transfer $\langle q_z \rangle $ that is approximately inverse
proportional to the number of scatterers $\langle n \rangle $, in
rough agreement with the signatures discussed above.

With regard to the threshold $n \ge 2$ for the occurrence of the
Bose-Einstein interference, the presence of a threshold implies a
sudden increase of the ridge yield as a function of centrality, as
represented by the number of participants.  Although the experimental
data with a high $p_T$ trigger appear to be consistent with the
presence of thresholds, the large error bars and the scarcity of the
number of data points in the threshold regions preclude a definitive
conclusion.

However, threshold effects for the ridge yield (2D Gaussian yield) as
a function of the number of participants, $N_{\rm part}$, have been
observed in another angular correlation measurements with a low-$p_T$
trigger from the STAR Collaboration \cite{Dau08,Ray08a,Ket09,STA11}.
We note previously that a fast jet parton possesses low-$p_T$ jet
fragments and a minimum-$p_T$-biased low-$p_T$ trigger can also
indicate the passage of a fast parent jet \cite{Won11}.  As a
consequence, ridge particles will also be associated with a low-$p_T$
trigger.  The sudden increase of the amplitude and the peak $\eta$
width of the ridge yield as a function of $N_{\rm part}$ may indicate
the presence of a threshold for the ridge yield as a function of
centrality \cite{Won11a}.

\section{Conclusions and Discussions}

~~~~In the collision of a fast particle with many medium particles,
the set of multiple collisions can occur incoherently as a sequence of
$n$ two-body collisions or coherently as a single (1+$n$)-body
collision.  From the binary collision data, one obtains the
longitudinal correlation length $\Delta z_{\rm coh}\sim 2\hbar
p_0/q_T^2$.  If the medium is so dilute, the binary cross section so
small, and the incident particle so slow that $\lambda \gg \Delta
z_{\rm coh}$, then the set of multiple collisions consists of a
sequence of incoherent collisions.  On the other hand, if the medium
is so dense, the binary cross section so large, and the incident
particle is so energetic such that $ \Delta z_{\rm coh} \gg \lambda $,
then the set of multiple collisions consists of a single coherent
(1+$n$)-body collision.

In central collisions of heavy nuclei at RHIC energies, the
environment of the produced medium favors the occurrence of coherent
collisions between the jet and medium particles.  The dynamics is
governed by Feynman diagrams linking the incident jet with the
scatterers as a connected unit.  For a set of initial states and a set
of final states in such a coherent collision, there are many different
trajectories in the sequences of collisions along $\Delta z_{\rm coh}$
at which various virtual bosons are exchanged.  By Bose-Einstein
symmetry, the total Feynman amplitude is the sum of all Feynman
amplitudes with all interchanges of the virtual boson vertices.

Remarkably, in high-energy collisions, the summation of these Feynman
amplitudes from different ways of ordering the virtual boson vertices
interfere with each other, resulting in the cancellation of some parts
of the Feynman amplitudes and leaving only the other parts of sharp
distributions.  The longitudinal momentum transfer to each scatterer
is then constrained to be the same as the energy transfer.  There is a
substantial flow of the longitudinal momentum from the jet to the
scatterers.  Furthermore, the transverse momentum distribution favors
small values of $q_T$.  As a consequence, the scatterers recoil
collectively along the jet direction, leading to a correlation of the
scatterers and the jet around $\Delta \phi \equiv \phi({\rm
  scatterer})-\phi({\rm trigger~jet}) \sim 0$.

For the coherent collision of an energetic parton with parton
scatterers in non-Abelian cases, we find that the complete
Bose-Einstein symmetry in the exchange of virtual gluons consists not
only of space-time exchange symmetry but also color index exchange
symmetry.  There is always a space-time symmetric and color-index
symmetric component of the Feynman amplitude that behaves in the same
way as the Feynman amplitude in the Abelian case, and the
corresponding recoiling partons behave in the same way as in the
Abelian case.  There is thus a finite probability for the parton
scatterers to emerge collectively along the incident trigger jet
direction, each carrying a significant fraction of the longitudinal
momentum of the incident jet.

The collective recoils will lead to the $\Delta \phi \sim 0$
correlation of the ridge particles with the high-$p_T$ trigger.  Such
a signature of the Bose-Einstein interference may have been observed
in the $\Delta \phi\sim 0$ correlation in the angular correlation
measurements of produced hadron pairs in central AuAu collisions at
RHIC \cite{Ada05}-\cite{Jia08qm}.  The centrality dependence of the
ridge yield in AuAu collisions at $\sqrt{s_{NN}}=200$ GeV with a
low-$p_T$ trigger \cite{Dau08,Ray08a,Ket09,STA11} may also be
consistent with the presence of a ridge threshold at $n=2$, as
expected in the quantum many-body effect of Bose-Einstein
interference.

The collective recoils of the scatterers from the Bose-Einstein
interference may be the origin of the the longitudinal momentum kick
along the jet direction postulated in the momentum kick model that has
been quite successful in the analysis of the angular correlations of
hadrons produced in high-energy heavy-ion collisions
\cite{Won07}-\cite{Won11}.  It is interesting to note that the present
Bose Einstein interference in coherent collisions is consistent with
previous results on the Bose-Einstein interference in the emission of
real photons and gluons in high-energy interactions and in the sum of
the ladder and cross-ladder loop diagrams in the collision of two
particles \cite{Che69}-\cite{Lam97b}.

\vspace*{0.6cm}
\noindent{\bf Acknowledgment}

The author would like to thank Profs. Vince Cianciolo, Horace
W. Crater, C. S. Lam, and Jin-Hee Yoon for helpful discussions. This
research was supported in part by the Division of Nuclear Physics,
U.S. Department of Energy.

\end{document}